\documentclass[aps,prl,twocolumn,showpacs,superscriptaddress]{revtex4-1}  

\usepackage{graphicx}  
\usepackage{dcolumn}   
\usepackage{bm}        
\usepackage{amssymb}   
\usepackage{amsmath}

\newcommand{\dif}{\partial}
\newcommand{\chiperp}{\chi_{\rm turb}}

\newcommand{\cg}{\kappa_c}
\newcommand{\width}{d_E}
\newcommand{\ky}{K_y}
\newcommand{\g}{\gamma_0}
\newcommand{\gnl}{\gamma_{\rm NL}}
\renewcommand{\mp}{\langle p \rangle}
\newcommand{\p}{\tilde p}
\newcommand{\pref}{p_{\rm ref}}

\renewcommand{\a}{\hat p}
\newcommand{\kx}{K_x}

\begin{document}

\title{$E \times B$ flow-induced shearing-merging of filaments: a Ginzburg-Landau model \\
of Edge Localized Mode cycles}
\author{M. Leconte}
\affiliation{Adv. Physics Research Div., National Fusion Research Institute, Daejeon, Korea}
\author{G.S. Yun}
\affiliation{Postech Physics Dept, Pohang, Korea and Asia Pacific Center for Theoretical Physics}
\author{Y.M. Jeon}
\affiliation{Boundary Physics Div., National Fusion Research Institute, Daejeon, Korea}

\begin{abstract}
We derive and study a simple 1D nonlinear model for Edge Localized Mode (ELM) cycles. The nonlinear dynamics of a resistive ballooning mode is modeled via a single nonlinear equation of the Ginzburg-Landau type with a radial frequency-gradient due to a prescribed $E \times B$ shear layer of finite extent. The nonlinearity is due to the feedback of the mode on the profile. We identify a novel mechanism, whereby the ELM only crosses the linear stability boundary once, and subsequently \emph{stays} in the nonlinear regime for the full duration of the cycles. This is made possible by the shearing and merging of filaments by the $E \times B$ flow, which forces the system to oscillate between a radially-uniform solution and a non-uniform solitary-wave like solution.
The model predicts a `phase-jump' correlated with the ELM bursts.
\end{abstract}

\maketitle

In magnetically-confined plasmas, edge localized modes (ELMs) \cite{Connor1998} are a virulent instability of
the plasma, responsible for a significant erosion of the plasma-facing components occurring in the high-confinement (H-mode) regime. Although they are beneficial for the expulsion of impurities from the plasma core, for future
devices such as ITER, they should be avoided. 
The present status of the theory of ELMs is the following: a linear ideal MHD mode is driven unstable by free-energy from the pressure-gradient $\nabla p$ - and also by edge currents $j$ -  the peeling-ballooning modes \cite{Dobrott1977, ConnorHastieWilson1998,Snyder2002}. The later evolution is less clear, as it is usually assumed that a certain `loop' in the $\nabla p$ v.s. $j$  stability diagram occurs, but this is a speculation with no strong foundation. Effects of uniform toroidal flow shear were studied by several authors, see e.g. Ref. \cite{Cooper1988,WaelbroeckChen1991} and shown to be stabilizing.
Moreover, there is a large body of experimental evidence and simulations pointing to the non-linear nature of the dynamics of ELM cycles and ELM burst \cite{Wenninger, Yun2012, KimLeePark2015, Krebs2013, Rhee2015}.
Ref. \cite{WilsonCowley2004} derived a model extending the linear ideal MHD theory perturbatively to the nonlinear regime, leading to faster-than-exponential growth. However, their analysis leads to a finite-time singularity, and hence cannot explain ELM cycles.
Single-fluid simulations of electrostatic turbulence showed periodic relaxations of a transport barrier in presence of a prescribed sheared flow \cite{Beyer2005}. \\
In this Letter, we present a low-dimensional model, based on the Ginzburg-Landau like coupling of an ideal ballooning-type mode to the background pressure gradient, taking into account the effect of a - prescribed - mean sheared flow with finite shear-layer width $\width \ll a$, where $a$ is the plasma minor radius.
The mechanism we propose is the following: a coherent pressure-driven mode - characterized by filaments along the magnetic field - reaches a saturated state by nonlinear coupling to the pressure profile. In presence of small cross-field turbulent diffusion, the mean sheared flow destabilizes this saturated state by changing the radial structure of the mode thereby inducing a shearing and merging of filaments, leading to quasi-periodic nonlinear oscillations. The key point is the \emph{synergy} between the nonlinearity and the sheared flow. Our proposed mechanism is primarely electrostatic, although electromagnetic effects - observed experimentally - may play a secondary role.
We consider nonlinear heat balance, which can be written:
\begin{eqnarray}
\frac{\dif}{\dif t} \delta p + V_E \frac{1}{r_s} \frac{\dif}{\dif \theta} \delta p = - \delta v_x  \frac{\dif  \mp}{\dif x} - \gamma_c \delta p
+ \chiperp \nabla_\perp^2 \delta p
\label{hb1}\quad  \\
\frac{\dif \mp}{\dif t} + \frac{\dif}{\dif x} \Big< \delta v_x \delta p \Big> =\chiperp \frac{\dif^2 \mp}{\dif x^2} +S
\label{hb2} \qquad
\end{eqnarray}
where $\nabla_\perp^2 = \dif_x^2 +r_s^{-1}\dif_\theta^2$ is the laplacian perpendicular to the magnetic-field, with $r_s$ the position of the mode resonance surface.
Here, the pressure is decomposed into an axisymmetric part $\mp$ and harmonics $\delta p = p - \mp$, where $\mp = \iint_0^{2 \pi} p dy dz$ is the poloidally and toroidally averaged pressure, and same for the velocity.
Magnetic shear effects - i.e. the fact that modes are localized in the vicinity of a resonance surface - are not taken into account in our model. Some effects of magnetic shear could be assessed in our model in the form of parallel heat diffusion, but this is beyond the scope of this Letter.
 
Note that the aim of this work is to study the basic dynamic process of ELM cycles, which can be understood from an electrostatic point of view. Therefore, in this simplified model, diamagnetic effects and magnetic fluctuations are not included, and we use the flute approximation ${\bf K}  \cdot {\bf B} \simeq 0$ - with $\bf K$ the wavenumber and ${\bf B}$ the magnetic field.
In Eqs. (\ref{hb1},\ref{hb2}) $\chiperp$ is a cross-field turbulent heat diffusivity, and $S=S(x)$ is an energy source modeling a constant heat-flux from the plasma core.
Following the standard convention, $x$ represents the local radial coordinate, $y=r_s \theta$ is the local poloidal coordinate, and $z=R \varphi$ is the local toroidal coordinate, in a magnetic fusion device.
The parameter $\gamma_c$ denotes the ideal MHD threshold. In order to obtain a transport barrier, the poloidally and toroidally averaged component ${\bf V}(x,t) = \frac{\dif \langle \phi \rangle}{\dif x} {\bf e}_y $ is prescribed in the form:
${\bf V}_E(x) = V_{E0}' \width \tanh(x / \width) {\bf e}_y $ corresponding to a shear-layer of finite extent chosen to be centered at $x=0$, where $V_{E0}'$ denotes the maximal shear rate and $\width =  (E_r' / E_r)^{-1}$ is the shear-layer width. Here, $\langle \phi \rangle = \iint_0^{2\pi} \phi dy dz$ is the poloidally and toroidally averaged electric potential. The $E \times B$ sheared flow can be related to toroidal sheared flow via $V_{tor} = B_p V_E$, with $B_p$ the poloidal magnetic field.
Considering the dominant toroidal harmonic $\delta p \sim \p \exp[-in(\varphi -q_0 \theta)] +c.c.$, where $q_0~(=m/n)$ is the safety factor at the mode location, we decompose the pressure and radial velocity complex amplitudes in the form
$ \p = |\p| \exp \Big(i \int \kx^p(x') dx' \Big) +c.c.$
and $\tilde v_x = |\tilde v_x| \exp \Big(i \int \kx^v(x') dx' \Big) +c.c. $
where $\kx^p, \kx^v$ denote the radial wavenumber associated to pressure and radial velocity, respectively.
We may then write the radial velocity as:
\[
\delta v_x = C_{vp} \p e^{i \delta_{vp}} e^{-in(\varphi -q_0 \theta)} +c.c.
\]
where $\delta_{vp} = \int (\kx^v - \kx^p) dx'$ denotes the cross-phase between radial velocity $\tilde v_x$ and pressure $\p$, and $C_{vp}$ is the ratio of amplitudes.

We assume $C_{v p}(x) = |\tilde v_x| / |\tilde p| $ and a negligeable cross-phase $\delta_{v p} \simeq 0$, leading to $\delta v_x \simeq C_{vp} \delta p$. This approximation is justified if the correlation-time of the cross-phase is large compared to the characteristic transport time-scale, the relevant regime for a coherent mode which we consider here. The opposite limit of small correlation-time was investigated recently \cite{XiXuDiamond2014}.
We obtain the following model coupling the complex amplitude $\p$ to $\mp$:
\begin{eqnarray}
\frac{\dif \p}{\dif t} +i\ky V_E(x)  \p & = & \g \left[ -\frac{\dif \mp}{\dif x}  - \cg \right] \p
+ \chiperp \frac{\dif^2 \p}{\dif x^2}
\label{harm1} \qquad \\
\frac{\dif \mp}{\dif t} & = & \chiperp \frac{\dif^2 \mp}{\dif x^2} - C_{v p} \frac{\dif}{\dif x} |\tilde p|^2+S(x)
\label{axi1}
\end{eqnarray}
where we used the approximation $|\dif / \dif x| \gg \ky$.
In Eq. (\ref{harm1}), $\ky=n q_0$ is the poloidal wavenumber, $\cg$ is the absolute-value of the critical gradient \emph{in absence of flow-shear}, related to the ideal MHD threshold. The second term on the r.h.s. of Eq. (\ref{axi1}) is due to the convective-flux $Q_{conv} = C_{v p}|\p|^2 \cos \delta_{v p} \simeq C_{vp} |\p|^2 $.
Hence, we will freely associate the mode internal energy $|\p|^2$ with the convective heat flux, both being quadratic in the spatial fluctuation fields.
Note that a similar assumption was made in the 0D model of Ref. \onlinecite{Bian2003}.
Since we expect the pressure gradient to be limited by the convective flux, we have $\dif \mp / \dif t \sim 0$, i.e. we use the following slaving approximation:
$- \frac{\dif}{\dif x} \mp \sim \frac{Q}{\chiperp} - \frac{C_{vp}}{\chiperp} |\p^2|$.
After some algebra, we obtain the following - normalized - Ginzburg-Landau type of equation:
\begin{equation}
\frac{\dif \p}{\dif t} +i \ky V_E(x) \p = \gamma_L(Q) \p + \frac{\dif^2 \p}{\dif x^2} -\gnl |\p|^2 \p
\label{gl0}
\end{equation}
The associated normalized pressure gradient is:
\begin{equation}
- \frac{\dif \mp}{\dif x} = \frac{Q}{Q_{c0}} - \gnl |\p^2|
\label{axi0}
\end{equation}
where $\gamma_L(Q) = Q / Q_{c0} - 1$ is the linear growth-rate, and $\gnl = (\pref/a) (C_{vp} / Q_{c0})$ represents the nonlinear damping. For simplicity, we take $C_{vp}$ as uniform in the numerical implementation.
Introducing the characteristic time $\tau = [(\g a / \pref) Q_{c0} / \chiperp]^{-1}$, time is normalized as $t / \tau \to t$. Here, $\pref$ is a reference pressure.
Following e.g. Ref. \onlinecite{Kuramoto2003}, space is normalized as $ x / \xi_{ref} \to x$, with $\xi_{ref} = \sqrt{\chi_{turb} \tau} $.
The analog of the Ginzburg-Landau correlation length is then $\xi = \xi_{ref} / \sqrt{\gamma_L(Q) \tau} $.
The normalized convective flux - the experimentally relevant quantity - is given by:
\begin{equation}
Q_{conv} = \frac{C_{vp}}{Q-Q_{c0}} |\p|^2
\label{defflux}
\end{equation}

Asymptotically far from the flow shear-layer $|x| \gg \width$,  the system (\ref{gl0}, \ref{axi0}) is bi-stable, with the two possible states I and IIa given in Table \ref{tab:table1}. The saturated state IIa, where the pressure profile `sits' near the linear threshold $Q_{c0}$ is a typical example of self-organized criticality (SOC). However, due to the finite shear-layer width, self-organized criticality \emph{breaks down} in the shear-layer, and non-linear oscillations set in.
We solved the nonlinear PDE (\ref{gl0}) numerically using a finite-difference scheme. The spatiotemporal dynamics - in presence of a mean sheared flow - of the mode squared amplitude $|\p|^2$ - proportional to the convective energy flux in our model - is shown [Fig. \ref{fig:fig2}a].
The system exhibits non-linear oscillations \emph{radially-localized} near the maximum of the flow-shear. The nonlinear oscillations share similarities with ELM bursts. The amplitude of the nonlinear oscillations tends to zero away from the shear-layer of extent $\width$.
Radial profiles of the heat-flux, are shown in the initial quiescent period time at $t=0$, and before a convective-burst at $t=80$ [Fig. \ref{fig:fig2}b]. The flux profile exhibits a soliton-like negative perturbation ($t=80$) of the reference profile ($t=0$). Phase-jumps, i.e. sudden changes in the eikonal $\alpha = \int \kx(x') dx'$ are correlated with the ELM bursts [Fig. \ref{fig:fig3}a].
The scaling law of pseudo-frequency  v.s. heat flux - in a limited parameter space - shows a decrease of burst frequency with heating power [Fig. \ref{fig:fig3}b]. The frequency was estimated using a peak-detection algorithm.

To understand the numerical results, further analytical progress can be obtained by using the ansatz: $ \p(x,t) = \a(x,t) e^{i \int \kx (x',t) dx' } $
which yields a system of coupled nonlinear PDE's known as Likharev equations in the framework of superconductivity [Eqs. 1a and 1b of Ref. \onlinecite{Likharev1976}]. Note that, in contrast to the usual ballooning approximation - we do not assume any a-priori scale separation between the enveloppe $\a$ and the wavenumber $\kx$. We stress here that keeping all the terms in the expansion is crucial for a nonlinear analysis, otherwise certain invariants of the system are lost. Nevertheless, rewriting the Ginzburg-Landau equation in terms of $\a$ and $\kx$ simplifies the analysis and allows to gain more physical insight.

In a very-wide shear-layer $\width \to +\infty$, the flow-shear is approximately uniform $V_E'(x) \simeq V'_{E0}$, and the amplitude $\a$ is approximately uniform $\a \simeq \a(t)$. It follows that the radial wavenumber varies linearly with time $\kx = - \ky V'_{E0} t$, this is the usual weak-shear regime. However, in a finite shear-layer $\width \ll a$, we obtain from Eq. (\ref{gl0})
 - after some algebra - the following nonlinear system of coupled PDEs:
\begin{eqnarray}
\frac{\dif \a}{\dif t} & = & \frac{\dif^2 \a}{\dif x^2} + \Big[ \gamma_L(Q) - \kx^2 \Big] \a -\gnl \a^3
\label{ode1} \\
\frac{\dif \kx}{\dif t} & = & \frac{\dif}{\dif x} \left[ \frac{1}{\a^2} \frac{\dif}{\dif x} \Big( \a^2 \kx \Big) \right] - \ky V_E'(x)
\label{ode2}
\end{eqnarray}
Eqs. (\ref{ode1},\ref{ode2}) have two limiting cases: i) without flow-shear $V_E'=0$, the second term on the r.h.s. of Eq. (\ref{ode2}) vanishes and the system admits a conserved quantity, the analog of the \emph{supercurrent} $\a^2 \kx = J = {\rm Cst}$. This allows a non-uniform saturated solution \cite{Langer1967}. The three possible states are given in Table \ref{tab:table1}. Apart from the trivial state $\p=0$, two states are possible, a uniform state and a non-uniform state characterized by a dark soliton, also known as \emph{caviton} - since it corresponds to a negative perturbation of the background - of the form:
\begin{equation}
u(x)= \frac{ \Delta }{ \cosh^2(x \sqrt{\Delta}) }
\label{caviton}
\end{equation}
, with $\Delta$ a real-valued parameter. ii) The linearized solution of Eqs. (\ref{ode1},\ref{ode2}) does not have a non-trivial steady-state ($\neq 0$) but shows a time-dependent shear-flow induced stabilization with a cubic exponential decay-rate $\tau_V^{-1} =  \chiperp^{1/3} [\ky V_E']^{2/3}$:
\[ \a_{lin} \sim e^{\gamma_L t - t^3/ \tau_V^3} \]
Clearly, when both the NL effects and the flow-shear effect are present, the system cannot have a (non-oscillating) stationary state. We hypothesize that this is the underlying reason for the bifurcation to a novel quasi-stationary state exhibiting NL oscillations. To confirm this interpretation, the 2D profile was reconstructed from the 1D complex amplitude, i.e.: $\delta p ={\rm Re} \{ \p \} \cos(\ky y) + {\rm Im} \{ \p \} \sin(\ky y)$. A series of snapshots reveals strong shearing and subsequent \emph{merging} of the filaments [Fig. \ref{fig:merging}]. The merging occurs at the same time of the ELM burst, which suggests that the filament merging causes the ELM burst.
We note that - to be more realistic - our model can be extended to include more toroidal harmonics, i.e. $n=n_1, n_2, \ldots$ nonlinearly coupled via the $n=0$ pressure profile. Additionally, the mean sheared flow should be self-consistently determined, i.e. from radial force balance.

\begin{figure}
\includegraphics[scale=0.25]{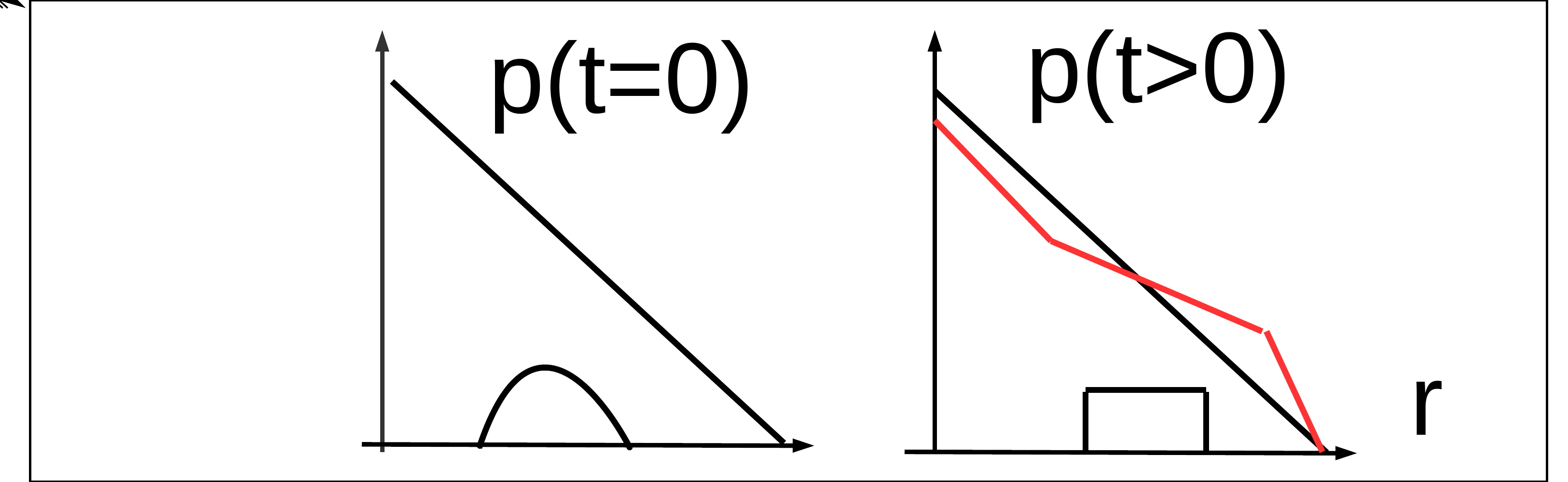}
\caption{\label{fig:sketch} Sketch of the Ginzburg-Landau model for ELMs. A mode grows at the center of the barrier, it saturates - by coupling to the profile -
and - in presence of a sheared-flow - subsequently undergoes nonlinear oscillations}
\end{figure}

\begin{figure}
\includegraphics[width=0.24\textwidth]{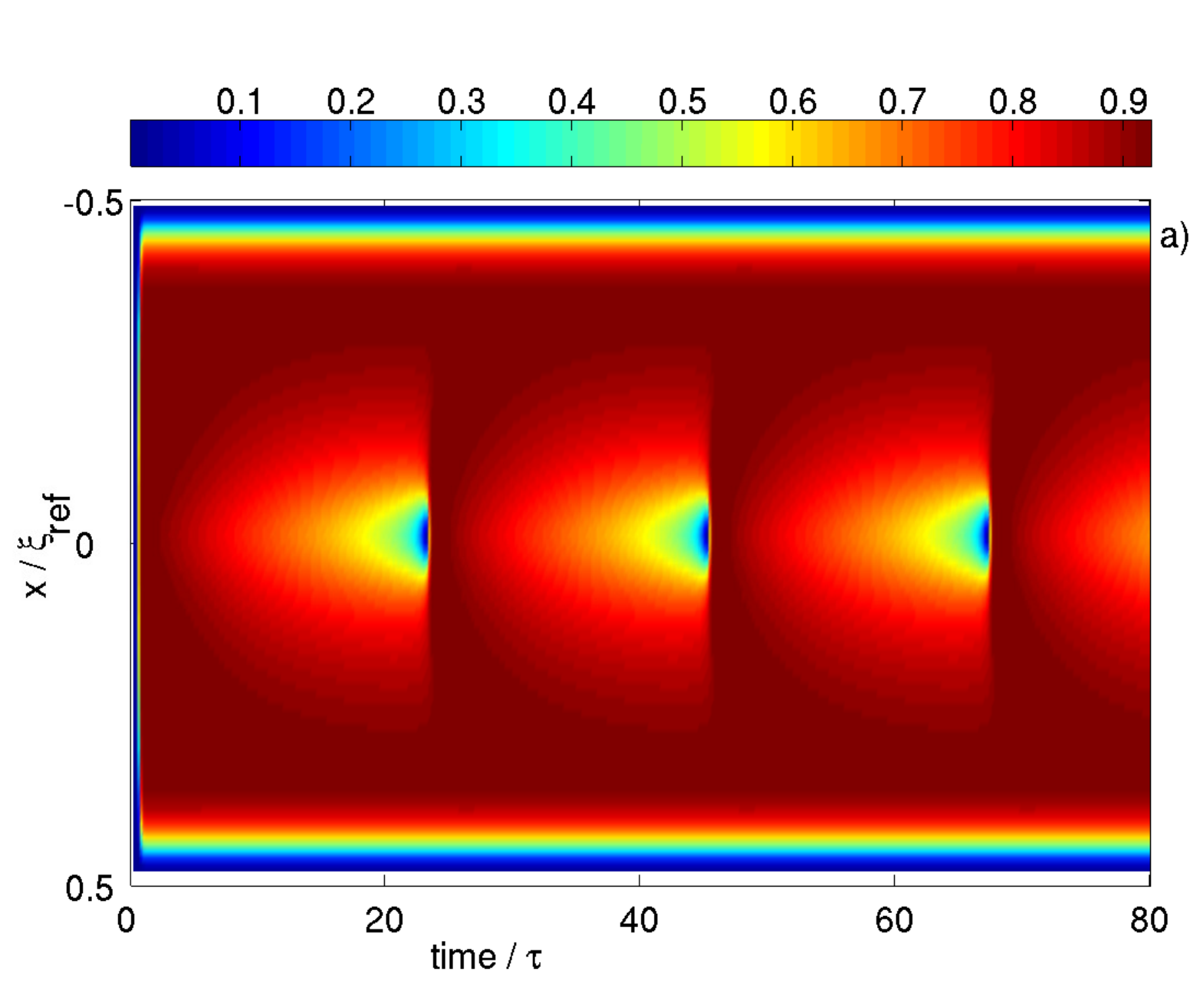}
\includegraphics[width=0.22\textwidth]{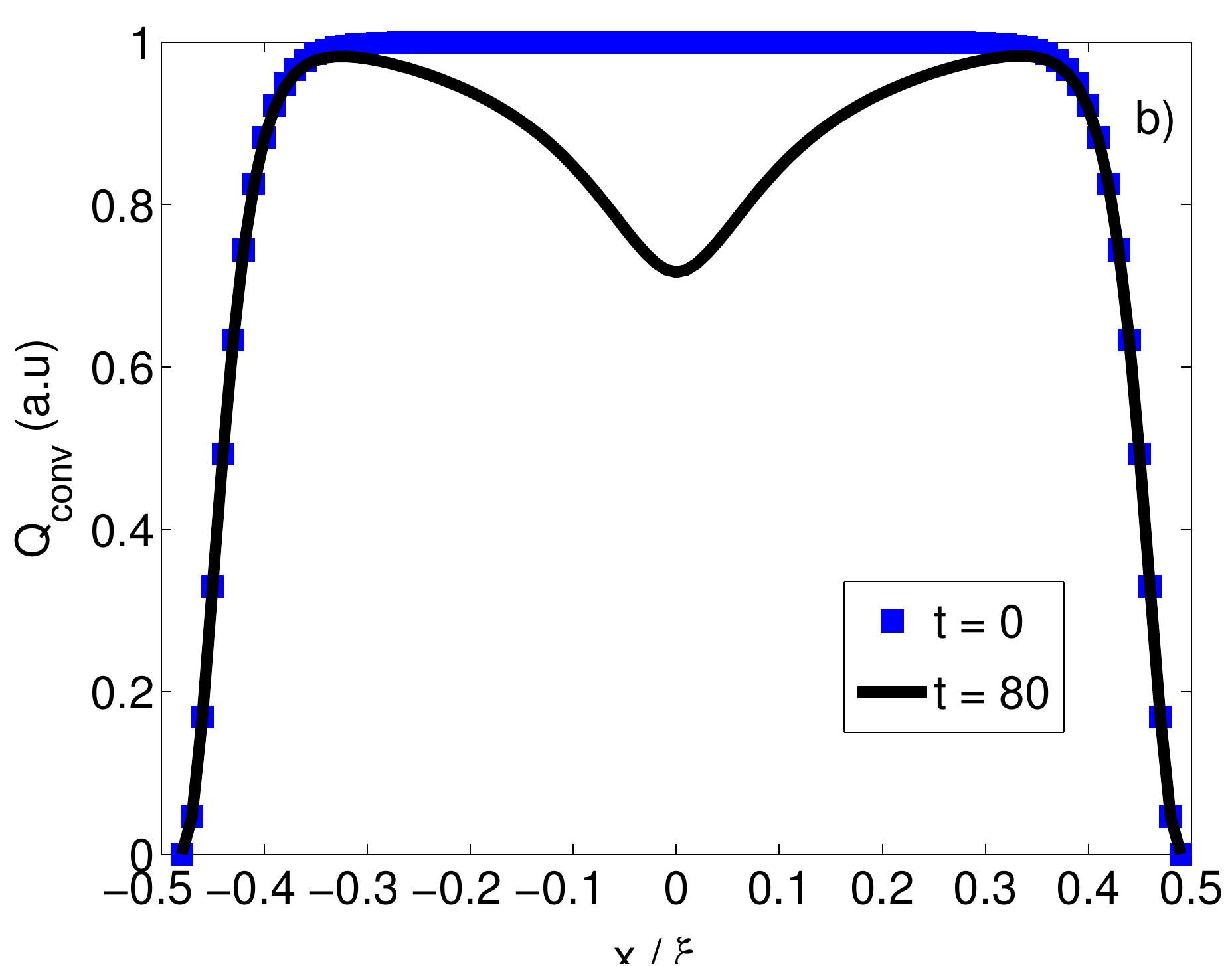} 
\caption{Spatiotemporal dynamics of the Ginzburg-Landau ELM model (\ref{gl0}), Panel a) shows an x-t contour plot of the instantaneous convective flux $Q_{conv}$, with a prescribed mean-sheared flow with maximal shear $V_{E0}' = 40$, and layer-width $\width=0.01$. Other parameters are: $Q/Q_{c0}=16.7$, $\ky=1$ and $\gnl=1$. Panel b) shows the radial profile of the instantaneous convectve flux, at time during the initial quiescent period ($t=0$) and before a heat burst ($t=80$).}
\label{fig:fig2}
\end{figure}

\begin{figure}
\includegraphics[width=0.25\textwidth]{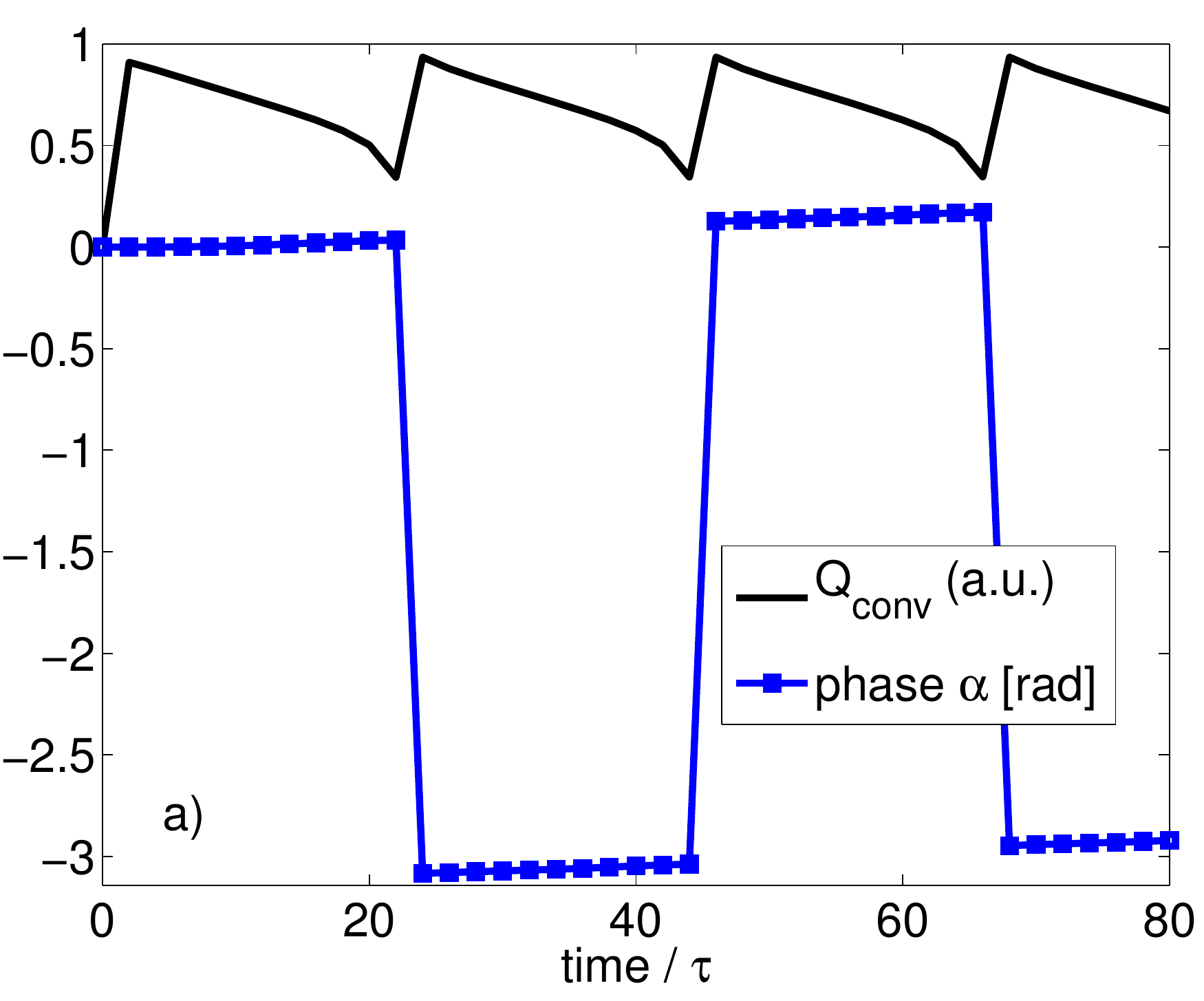} \\
\includegraphics[scale=0.25]{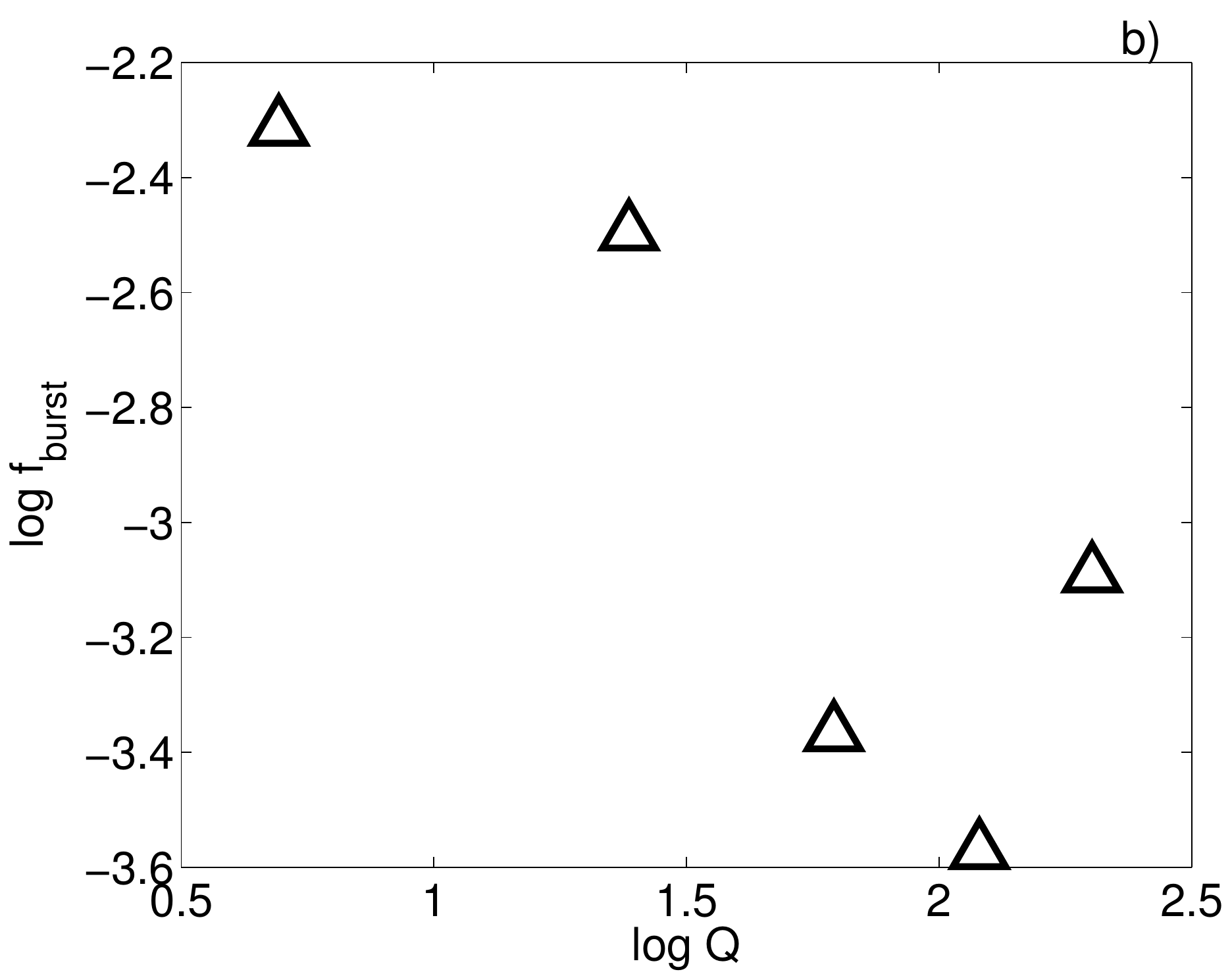}
\caption{Phase jumps are correlated with convective bursts, Panel a) shows timeseries of the convective flux (solid-line) at the resonance surface $x=0$, and the associated jumps in the phase i.e. eikonal (squares).  Parameters are the same as in Fig. \ref{fig:fig2}. Panel b) shows the scaling-law of heat burst frequency v.s. total heat-flux, i.e. heating power, $Q$.}
\label{fig:fig3}
\end{figure}

\begin{figure}
\includegraphics[scale=0.25]{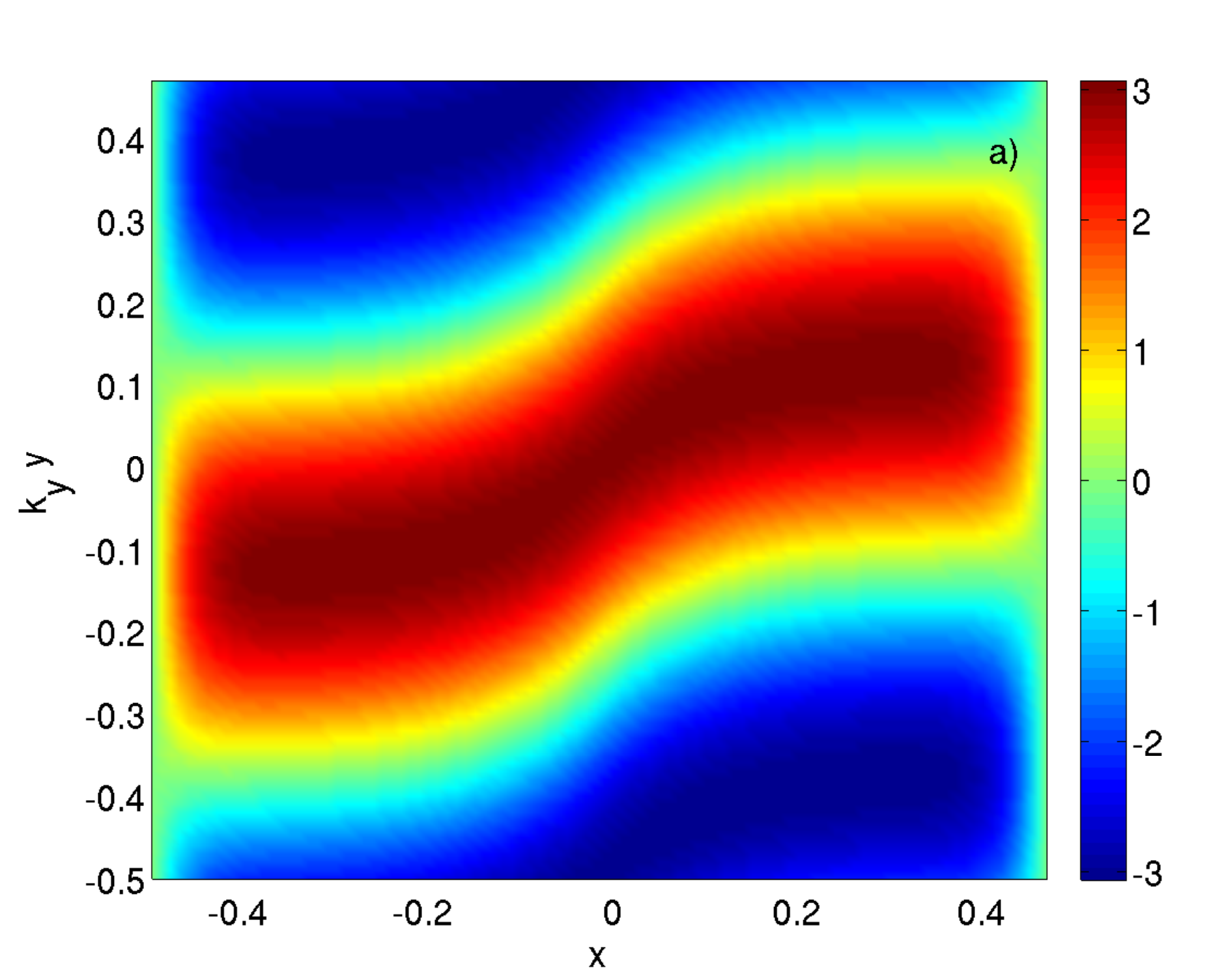}\includegraphics[scale=0.25]{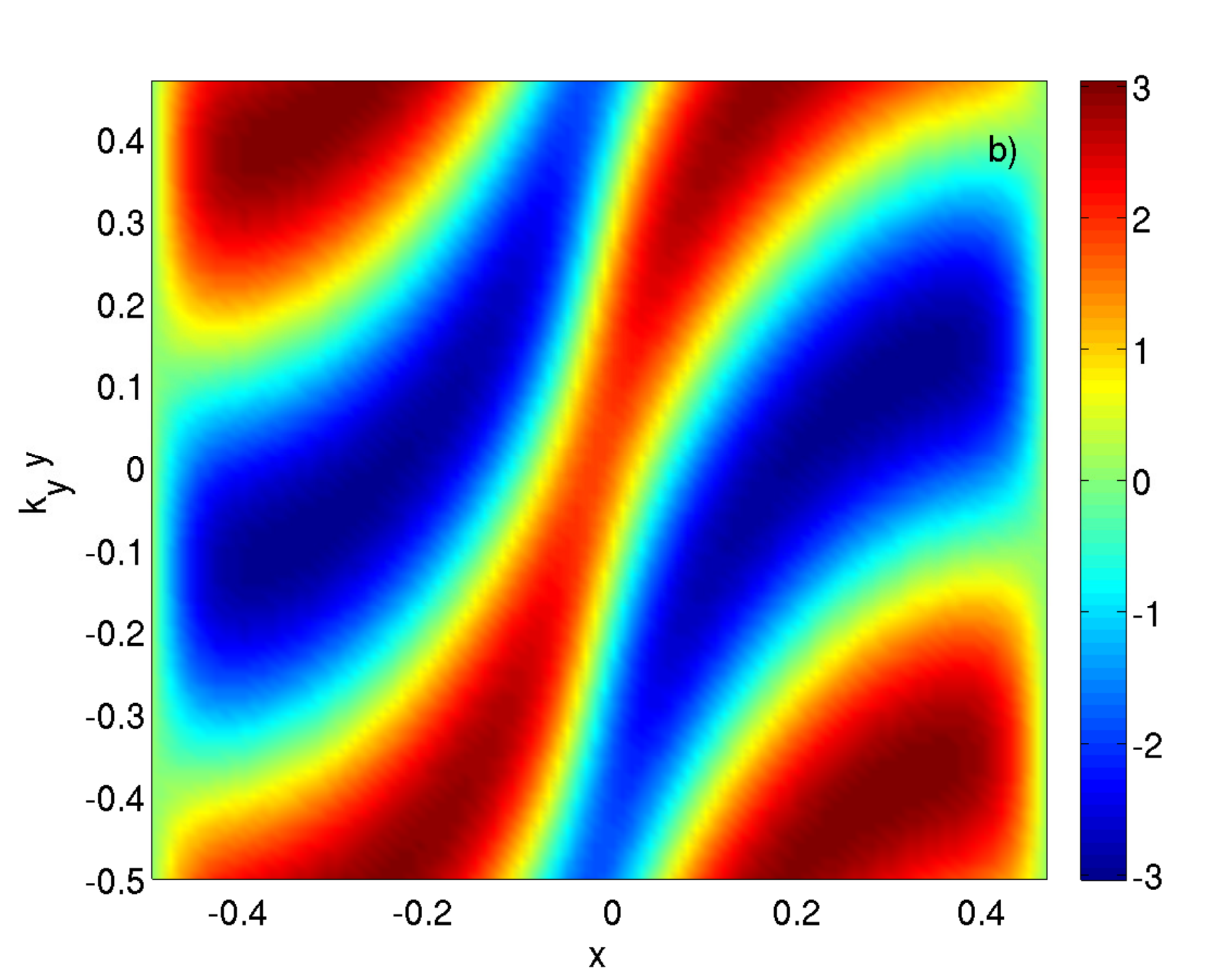}\\
\includegraphics[scale=0.25]{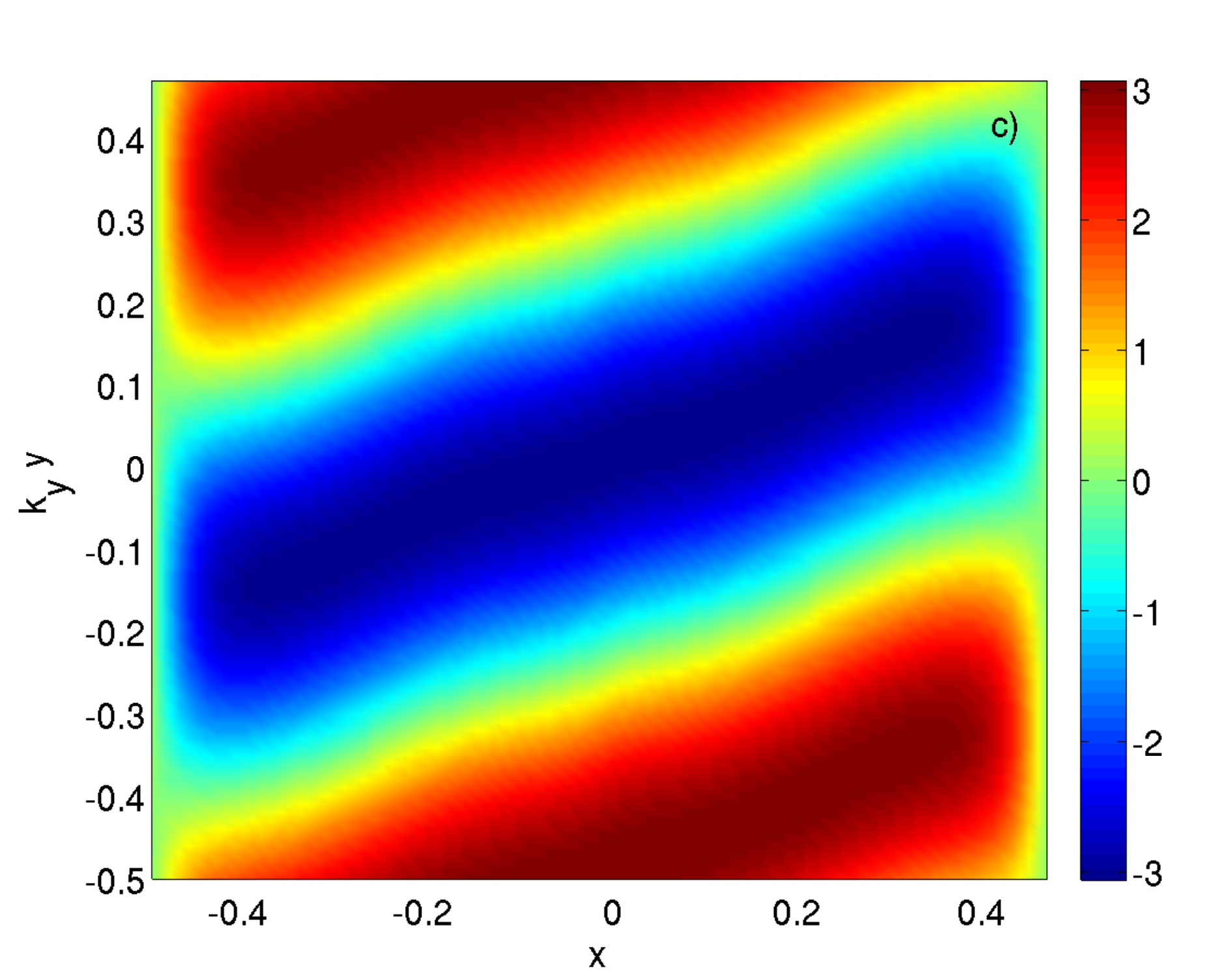}\includegraphics[scale=0.25]{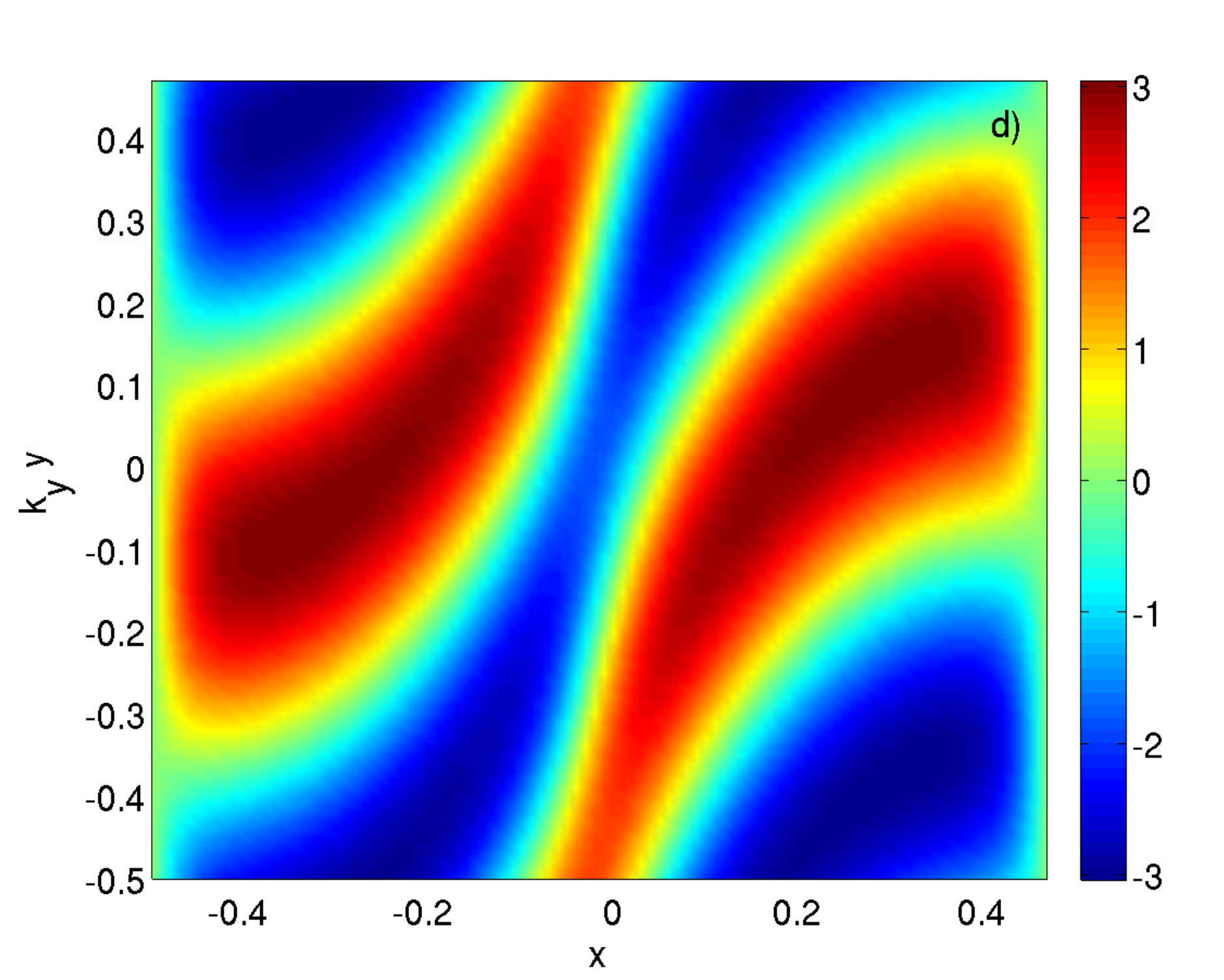}
\caption{Shearing and merging of filaments in a shear-layer of finite extent ($\width=0.01$). Snapshots of the mode $\delta p$ reconstructed from the 1D complex amplitude. Snapshots are taken at time: a) t=4, b) t=24, c) t=26 and d) t=46. Parameters are the same as in Fig. \ref{fig:fig2}. Filament merging - correlated with a heat burst - occurs in-between b) and c).}
\label{fig:merging}
\end{figure}

\begin{table}
\caption{\label{tab:table1} Saturated states of the reference system without flow-shear $V_{E0}'=0$. The function $u(x)$ is given by Eq. (\ref{caviton}). }
\begin{ruledtabular}
\begin{tabular}{l|c|r}
 & $|\p|(x)$ & $- \dif_x \mp$ \\
\hline
state I & 0 & $\frac{Q}{Q_{c0}}$ \\
state IIa & $\sqrt{ \gnl^{-1} \left[ \frac{Q}{Q_{c0}}-1 \right] }$ & $1$ \\
state IIb & $\sqrt{ \gnl^{-1} \left[ \frac{Q}{Q_{c0}}-1 \right] } [1-u(x)]^{1/2}$ & $1+\left[ \frac{Q}{Q_{c0}}-1 \right]u(x)$
\end{tabular}
\end{ruledtabular}
\end{table}

In conclusion, we derived and studied a simple 1D nonlinear model for ELM cycles. Our numerical results and analysis provide a novel mechanism, whereby the ELM only crosses the linear stability boundary once, and subsequently \emph{stays} in the nonlinear regime for the full duration of the cycles. This is made possible by the shearing and merging of filaments by the $E \times B$ flow, which forces the system to oscillate between a uniform solution and a non-uniform solution. We caution, however, that this finding applies only to the purely pressure-driven case.

\section*{Acknowledgements}
One of the authors (M. Leconte) would like to thank P. Beyer and X. Garbet and the participants of the 2013 'Festival de Theorie, Aix en Provence'  for usefull discussions.
This work was supported by R\&D Program through National Fusion Research Institute (NFRI) funded by the Ministry of Science, ICT and Future Planning of the Republic of Korea (NFRI-EN1541-1).

\end{document}